\documentclass[aps,prl,reprint]{revtex4-1} 
\usepackage{amsmath}
\usepackage{graphicx}
\usepackage{dcolumn}
\usepackage{multirow}
\usepackage[flushleft]{threeparttable}

\pdfmapfile{=fullembed.map}

\renewcommand{\Re}{\operatorname{Re}}
\renewcommand{\Im}{\operatorname{Im}}

\begin{document}


\title{Towards high cooperativity strong coupling of a quantum dot in a tunable microcavity}



\author{Lukas Greuter}
\author{Sebastian Starosielec}
\author{Andreas V. Kuhlmann}
\author{Richard J. Warburton}
\affiliation{Department of Physics, University of Basel, Klingelbergstrasse 82, Basel 4056, Switzerland}


\date{\today}

\begin{abstract}


We investigate the strong coupling regime of a self-assembled quantum dot in a tunable microcavity with dark-field laser spectroscopy. The high quality of the spectra allows the lineshapes to be analyzed revealing subtle quantum interferences. Agreement with a model calculation is achieved only by including exciton dephasing which reduces the cooperativity from a bare value of 9.0 to the time-averaged value 5.5.  In the pursuit of high cooperativity, besides a high-Q and low mode-volume cavity, we demonstrate that equal efforts need to be taken towards lifetime-limited emitter linewidths.
\end{abstract}

\pacs{}

\maketitle 

Cavity quantum electrodynamics (QED) involves an exchange of energy quanta between a single emitter and a cavity photon. The coupling rate $\hbar g = \mu_{12} E_\text{vac}$, depending on the emitter's dipole moment $\mu_{12}$ and the vacuum electric field at the location of the emitter $E_\text{vac}$, sets the relevant timescale of the coupled dynamics. If $g$ is considerably smaller than the emitter relaxation rate $\gamma$ or the cavity photon decay rate $\kappa$, on resonance the cavity mode acts as an additional decay channel to the emitter giving rise to an enhanced spontaneous emission rate (the Purcell effect of the weak coupling regime). If $g$ is much larger than the energy loss rates, a coherent exchange of energy quanta takes place giving rise to new eigenstates, ``polaritons", split in energy by $2 \hbar g$ (the strong coupling regime). The efficacy of the coherent coupling is commonly denoted by the cooperativity parameter $C = 2 g^2 / (\kappa \gamma)$, the figure of merit for this work. The coherent exchange was first realized with single Cs atoms in a high finesse cavity~\cite{BocaPRL2004}. 

The strong coupling regime is a potentially powerful tool in quantum information processing~\cite{ImamogluPRL1999}, notably in quantum networks~\cite{KimbleNature2008}, since it enables for instance atom-atom entanglement~\cite{PlenioPRA1999} or the distribution of quantum states~\cite{WilkScience2007}. Furthermore, strong coupling enables a nonlinear photon-photon interaction and hence the observation of photon blockade~\cite{BirnbaumNature2005,Reinhard2012}, a prerequisite for the creation of a single photon transistor~\cite{ChangNatPhys2007,VolzNature2012}. 

It is clearly desirable to implement cavity-QED in the solid-state as the solid-state host acts as a natural trap for the emitter. Furthermore, on-chip integration of multiple elements is feasible. As emitter, self-assembled quantum dots have desirable properties: high oscillator strength, narrow linewidths and weak phonon coupling~\cite{WarburtonNatMat2013}. As host, a semiconductor such as GaAs is very versatile: heterostructures can be realized; there is a wide array of post-growth processing techniques. Photoluminescence experiments on single InGaAs SAQD coupled to a photonic crystal cavity or a micropillar cavity revealed an anticrossing, the signature of the strong coupling regime~\cite{ReithmaierNature2004,YoshieNature2004,HennessyNature2007}. For micropillars, recent experiments exhibit cooperativity values of around $C \simeq 3$~\cite{LermerPRL2012}. For photonic crystal cavities, a much higher $C$ is achieved~\cite{EnglundNature2007} but $C$ is skewed by the fact that $g \gg \gamma$ yet $g \gtrsim \kappa$. The photon decay rate $\kappa$ at the emitter wavelength is relatively high in both geometries, limiting the cooperativity. In addition, micropillars and photonic crystals offer only limited spectral tuning to the emitter transition, and spatial positioning of the emitter relative to the cavity antinode is achieved either by good fortune or by fabricating the cavity around a particular emitter~\cite{Badolato2005,Dousse2008}. These are challenging issues resulting in a low yield.   

In this work we demonstrate a strong coupling of a single self-assembled InGaAs quantum dot to a fully tunable, miniaturized Fabry-Perot cavity~\cite{BarbourJAP2011,GreuterAPL2014}. The coupled emitter-cavity system is investigated by dark-field laser spectroscopy, yielding extremely high spectral resolution, high sensitivity, a high contrast and good mode-matching. The strong coupling regime is accessed definitively: we reach a cooperativity of $C=5.5$, significantly larger than that achieved with micropillars~\cite{LermerPRL2012} or a fibre-cavity~\cite{MiguelNJP2013}. The high quality of the data allows for a quantitative lineshape analysis. We demonstrate an interference in the polariton gap. However, the interference is less pronounced than expected from the ``standard model", the Jaynes-Cummings Hamiltonian. We show that the missing interference arises as a consequence of an additional emitter broadening. Including the emitter broadening allows us to reproduce both the exact lineshapes and polariton eigenenergies with a single parameter set for all cavity-emitter detunings. A key point emerges. Achieving a high cooperativity requires more than a focus on the cavity properties (small mode volume and high $Q$-factor): this has to be matched with an equal effort on improving the linewidth of the emitter. Here, we show that suppressing the emitter broadening would yield a cooperativity as high as $C = 9.0$ even with the present microcavity. Characterization of the quantum dots shows that here the main emitter broadening arises from a spectral fluctuation (rather than a true dephasing process): the fluctuations can be circumvented in lower-noise devices. Our system therefore represents an extremely promising route to implementing cavity-QED in the solid-state.


\begin{figure}
\includegraphics[width=85mm]{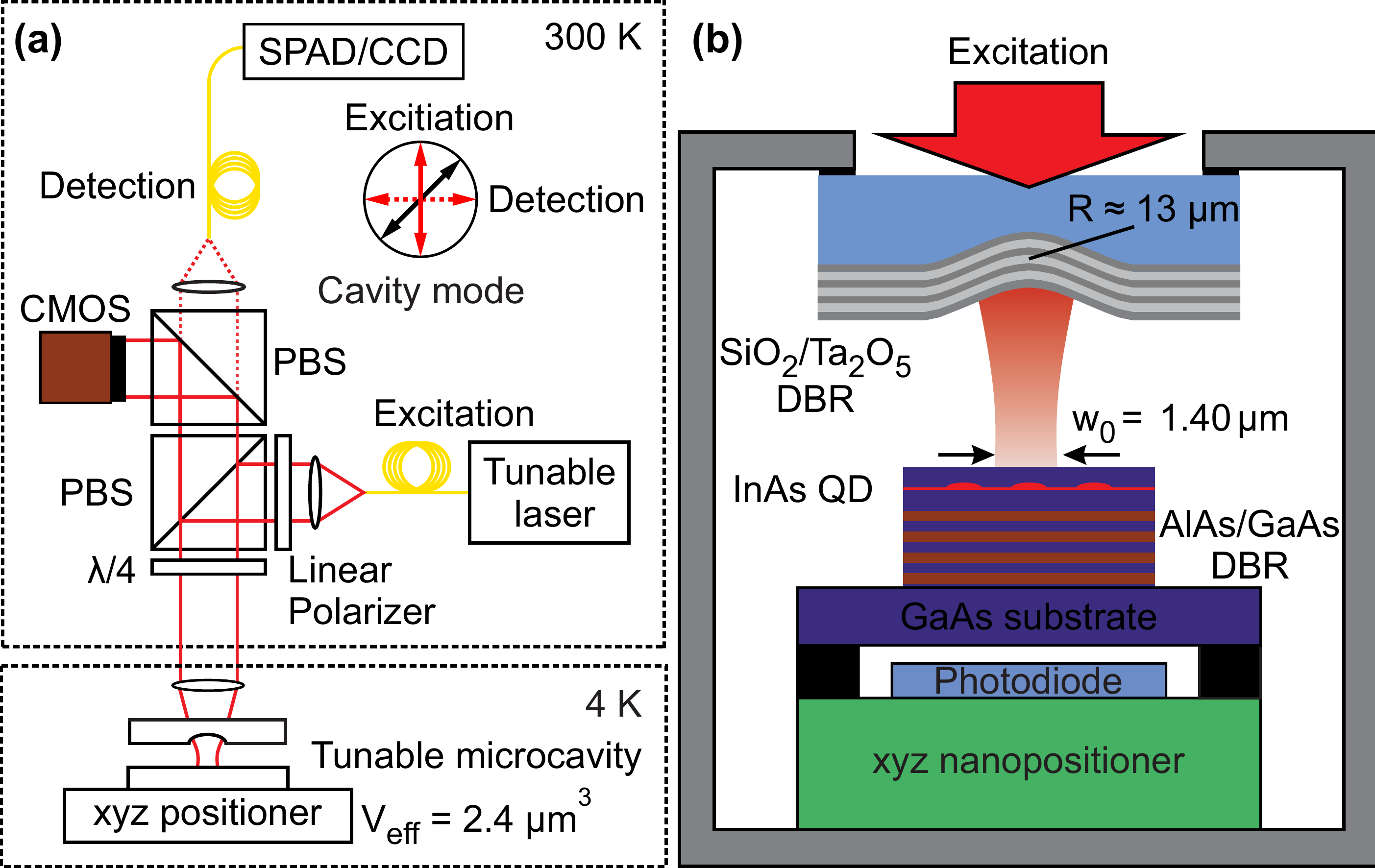} 
\caption{
\label{fig:setup}
(a) Experimental setup to probe the cavity-quantum dot system. The microscope head is at room temperature and consists of two polarizing beam splitters (PBS), a linear polarizer and a $\lambda/4$ waveplate. (b) Microcavity setup with a
GaAs/AlGaAs bottom mirror and a curved top mirror with radius of 13 $\mu$m coated with $\text{Ta}_2\text{O}_5/\text{SiO}_2$ DBR. The InGaAs quantum dots are embedded in a semiconductor heterostructure, at distance $\lambda/2$ from the surface and $\lambda/2$ from the bottom mirror.
}
\end{figure}

The emitter is a self-assembled InGaAs quantum dot grown by MBE at UCSB California. The details of the heterostructure are depicted in figure~\ref{fig:setup}b: a 32.5 pair $\lambda/4$ AlGaAs/GaAs distributed Bragg reflector (DBR) is terminated by a $\lambda$ layer of GaAs, which incorporates the InGaAs quantum dots in the center. 
The bottom DBR with reflectivity $R_\text{bot}=99.99\,\%$ forms the planar end mirror of the cavity. The concave top mirror consists of a fused silica substrate with a depression formed by $\text{CO}_2$ laser ablation~\cite{HungerAIPAdv2012}, and is coated with a $\text{Ta}_2\text{O}_5/\text{SiO}_2$ DBR of reflectivity $R_\text{top} = 99.95\,\%$. The radius of curvature is approximately $13\,\mu\text{m}$. The bottom semiconductor sample is mounted on an xyz piezo stack that allows for sub-nm positioning with respect to the top mirror enabling both spectral and spatial tuning. The whole microcavity is then mounted on another xyz piezo stack that allows the microcavity to be positioned with respect to an aspherical coupling lens ($\text{NA} = 0.55$), facilitating efficient mode matching with the excitation beam. A Si-photodiode mounted underneath the bottom mirror is used for transmission measurements to characterize and optimize the mode matching. 
By determining the longitudinal mode index $q_0 = 2 \partial L / \partial \lambda = 18$, we estimate an effective cavity length of $L = q_0 \lambda/2 = 8.5\,\mu\text{m}$. From these parameters, a Gaussian optics estimate results in a beam waist of $w_0 = 1.4\,\mu\text{m}$ at the sample. The cavity finesse is 4,000; the quality factor is $Q = 6 \times 10^4$.

We measure the coupled cavity-quantum dot dynamics with confocal cross-polarized dark-field laser spectroscopy~\cite{KuhlmannRSI2013}, sketched in figure~\ref{fig:setup}a.
The polarizing beam splitters (PBS) define two orthogonal linearly-polarized arms (excitation and detection) each coupled to the microcavity via the same objective lens. A linear polarizer and a quarter-wave plate mounted on piezo-driven rotational stages compensate for small imperfections in the optics and enable a suppression of the excitation laser of $10^{-7}$ to be reached, stable over several days. 
The cavity exhibits non-degenerate linearly-polarized longitudinal modes with a splitting of about $200\,\mu\text{eV}$, conveniently larger than the bandwidth required to probe fully the dynamics of the strong coupling. The cavity modes are aligned with respect to the polarization axis of the microscope at an angle $\phi \approx \pi/4$ allowing a good coupling of the cavity mode to both detection and excitation channels. We measure the wavelength of the tunable excitation laser with a wavemeter and use this information to calibrate the cavity detuning on applying a voltage to the microcavity z-piezo. While the polarization optics are all at room temperature, the microcavity setup is inserted into a stainless steel tube containing He exchange gas and cooled to $4\,\text{K}$ in a He bath cryostat.

\begin{figure}
\includegraphics[width=85mm]{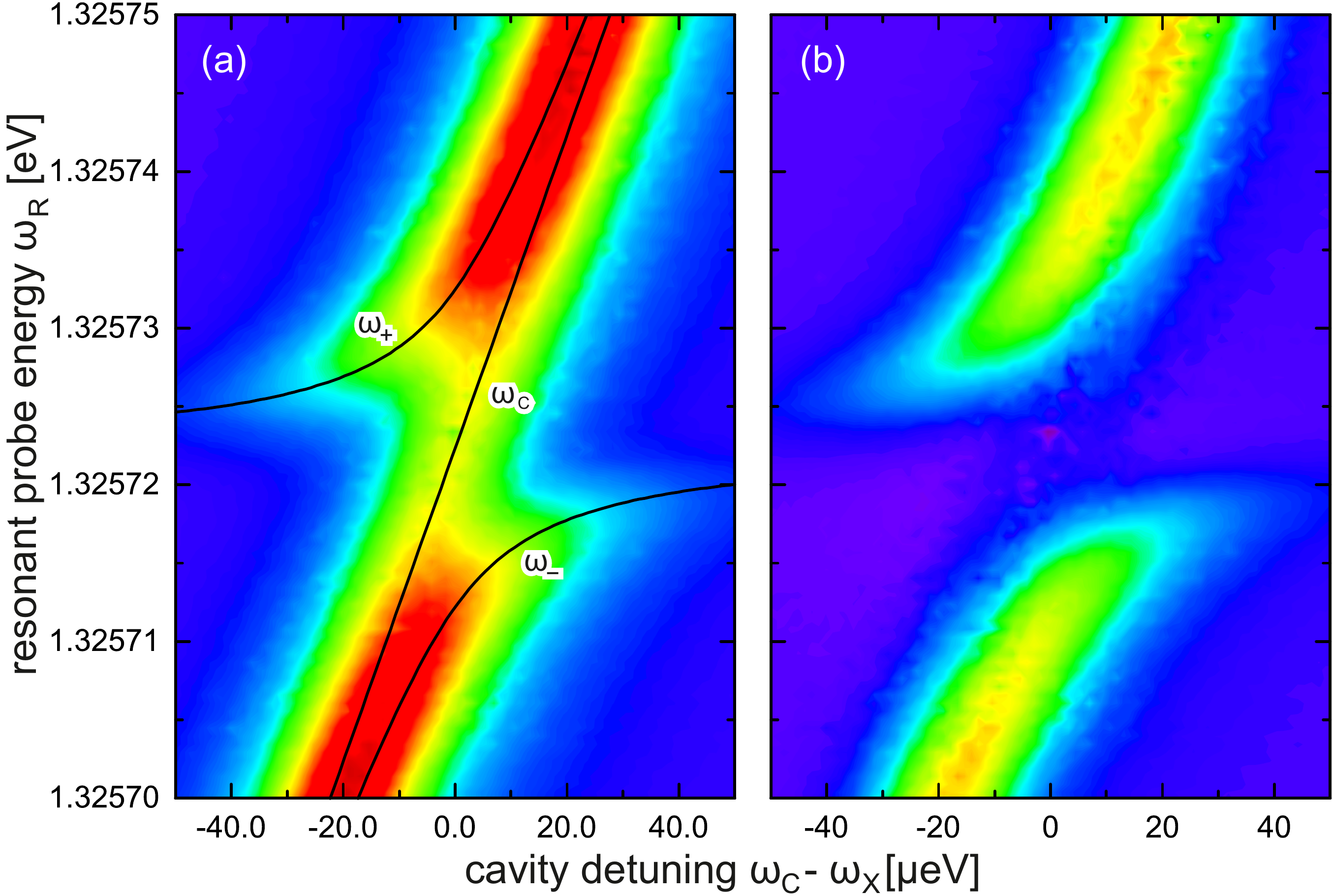} 
\caption{
\label{fig:rawdata}
Dark-field resonant laser spectroscopy on a coupled quantum dot-cavity system for varying cavity detuning. (a) A triplet is observed at resonances $\omega_\text{R} = \omega_\pm, \omega_\text{C}$. We interpret the spurious (bare) cavity resonance $\omega_\text{C}$ as a consequence of an unstable emitter state resulting in telegraph-like dynamics. (b) The data in (a) after subtracting the bare cavity resonance revealing the normal mode splitting characteristic of the strong coupling regime.
}
\end{figure}

Tuning the microcavity resonance with respect to the emitter transition, and sweeping the excitation frequency with respect to the microcavity resonance, reveals the exact lineshape of the coupled emitter-cavity system for various detunings, as shown in figure~\ref{fig:rawdata}a. We observe a triplet structure featuring the bare cavity resonance $\omega_\text{C}$ along with two detuning-depending resonances identified as the polariton states of the strong coupling regime. The bare cavity contribution can be determined accurately from the data in the polariton gap at zero detuning. A subtraction of the bare cavity resonance from the raw data reveals the clear anticrossing of the polariton modes, figure~\ref{fig:rawdata}b.

The anticrossing feature figure~\ref{fig:rawdata} is visible only if the sample is illuminated with an additional ultraweak non-resonant excitation laser ($\lambda = 830\,\text{nm}$). In free space laser spectroscopy experiments on a sample from the same MBE, an ``optical gating" by weak non-resonant excitation is described~\cite{NguyenPRL2012}. However, it is only partially successful: observation of the bare-cavity mode shows that the quantum dot detunes abruptly (and out of resonance with the microcavity) in a telegraph fashion. A bare-cavity contribution to resonance spectra has been observed also on photonic crystal cavities~\cite{HennessyNature2007} and was attributed to charge noise in the vicinity of the quantum dot, a mechanism which is active here. The experiment integrates over a much longer timescale than is typical for this telegraph noise, thus capturing photons from the scattering off the bare cavity a significant fraction of time. We do not observe a fine structure splitting of the exciton at zero magnetic field. A neutral exciton without fine structure is unlikely for these quantum dots~\cite{SeidlPhysicaE2007} so that we can safely assume that the studied exciton coupling to the cavity in figure~\ref{fig:rawdata} is a charged exciton.  

We model the experiment with the Jaynes-Cummings Hamiltonian modified for coherent excitation at frequency $\omega_\text{R}$
\begin{equation}
\mathcal{H} = \hbar \omega_\text{C} \, a^\dagger a + \hbar \omega_\text{X} \, b^\dagger b + [\hbar g \, a^\dagger b  +  \hbar \epsilon \, a^\dagger \text{e}^{-\text{i}\omega_\text{R} t} + \text{h.c.}]\;,
\end{equation}
Here, $a$ ($b$) is the bosonic (fermionic) annihilation operator of the microcavity photon (exciton transition) with energy $\hbar \omega_\text{C}$ ($\hbar \omega_\text{X}$); $g$ denotes the coherent coupling rate between photon and exciton; and $\epsilon$ is the effective coupling rate from the resonant excitation to the cavity field. Losses in the system are described by the Lindblad formalism including the photon energy loss rate $\kappa$ and the exciton relaxation rate $\gamma$. 
The cavity emission is modeled to be weakly coupled to a continuum of detection modes with overall collection efficiency $\eta$: the detected count rate is thus $\dot{N} = \eta \kappa \langle a^\dagger a \rangle$.

With model M1 we investigate the system's response as a function of the resonant probe frequency $\omega_\text{R}$, treating $\epsilon$ as a perturbative parameter. The linear coupling gives rise to two polariton modes ($\pm$) at Rabi frequencies $\omega_\pm$. The steady-state cavity population (proportional to the photon count rate) evaluates to
\begin{equation}
\begin{split}
\label{eq:decomposition}
\langle a^\dagger a \rangle(\omega_\text{R}) &= A^\text{L}_{-} \mathcal{L}(\omega_\text{R} - \omega_{-}) + A^\text{L}_{+} \mathcal{L}(\omega_\text{R} - \omega_{+}) \\ 
&+ A^\text{D} \mathcal{D}(\omega_\text{R} - \omega_{+}) - A^\text{D} \mathcal{D}(\omega_\text{R} - \omega_{-}) \;,
\end{split}
\end{equation}
where $\mathcal{L}(\omega) = \Im [(\pi \omega)^{-1}]$ is the unit-area Lorentzian function, 
$\mathcal{D}(\omega) = \Re [(\pi \omega)^{-1}]$ its dispersive function counterpart, each with peak location $\Re \omega = 0$ and FWHM parameter $2 \left| \Im \omega \right|$.
The peak areas $A^\text{L}_\pm$, $A^\text{D}$ and Rabi frequencies $\omega_\pm$ are closed form functions of the dynamical parameters $(g,\kappa,\gamma,\epsilon)$ (see Supplemental Material).

\begin{figure}
\includegraphics[width=85mm]{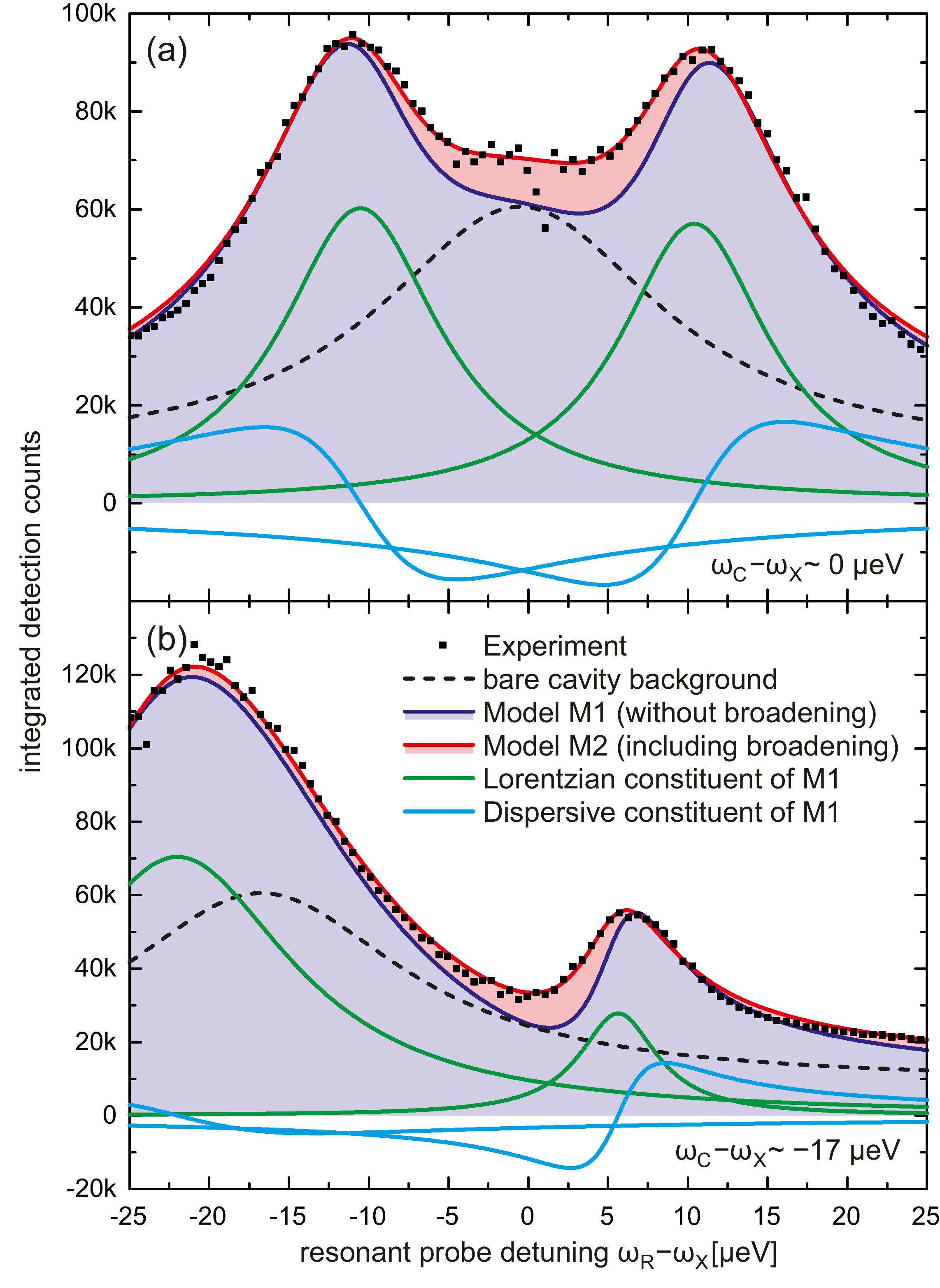} 
\caption{
\label{fig:resonance}
Dark-field laser spectroscopy: spectra for (a) zero and (b) $-17\,\mu\text{eV}$ cavity-emitter detuning. The experimental values (black dots) are globally fitted to model M1 (purple solid line), with Lorentzian and dispersive constituents (green and blue solid line), and to model M2 (red solid line). M2, which includes an additional broadening mechanism of the emitter, describes the experimental data much better than M1. The improvement is partially masked by the bare cavity resonance background (black dashed line).
}
\end{figure}

Figure~\ref{fig:resonance} shows (black dots) two exemplary lineshapes, (a) for zero cavity--exciton detuning $\omega_\text{C} - \omega_\text{X} = 0\,\mu\text{eV}$, and (b) for significant detuning $\omega_\text{C} - \omega_\text{X} = -17\,\mu\text{eV}$. The purple solid line shows a best $\chi^2$ fit of the observed counts to the model M1 eq.~\eqref{eq:decomposition}, where the fit results in a single set of dynamical parameters ($g, \kappa, \gamma, \epsilon$), a set used for all employed detunings (table~\ref{tab:fit}). The green and blue solid lines show the Lorentzian and dispersive constituents of the model, while the black dashed line represents the spurious bare-cavity contribution. The dynamical parameters obtained from the fit result in a cooperativity of $C = 2 g^2/(\kappa \gamma) = 5.5 \pm 0.1$.

Qualitatively, the model M1 agrees well with the observed polariton resonances in terms of splitting, linewidths as well as their shift with caviy--exciton detuning. Quantitatively however, the count rates within the polariton gap are significantly underestimated with respect to the experimental data for all detunings. In the polariton gap, the model (neglecting of course the bare-cavity contribution) predicts a strong destructive interference: the positive Lorentzian contributions are reduced considerably by the two dispersive constituents, both of which turn negative. In the experiment, this interference is observed to a lesser degree than that predicted by model M1. This lack of interference is particularly prominent for large detunings at the exciton-like polariton resonance (figure~\ref{fig:resonance}b) and points strongly to an emitter dynamic not considered by the model.

To investigate this missing dynamic, we performed independent linewidth measurements on the same sample region but without the top mirror. The linewidths are measured under the same conditions, i.e.\ with resonant laser spectroscopy in the presence of an ultraweak non-resonant excitation (see Supplemental Material). The results demonstrate a significant contribution to the exciton linewidth beyond that determined by spontaneous emission: typical linewidths are $3-4\,\mu\text{eV}$; the radiative-lifetime limited  linewidth (the ``transform limit") corresponds to $0.8\,\mu\text{eV}$. There are two culprits for this additional broadening: a spectral fluctuation (i.e.\ a wandering of the exciton central frequency on timescales longer than the radiative decay time) and pure exciton dephasing. The analysis (see Supplemental Material) suggests spectral fluctuations are dominant, but the exact conclusion is quantum dot dependent. Linewidth broadening on this scale is commonly observed and arises from electric charge noise~\cite{HouelPRL2012}. 

\begin{figure}
\includegraphics[width=65mm]{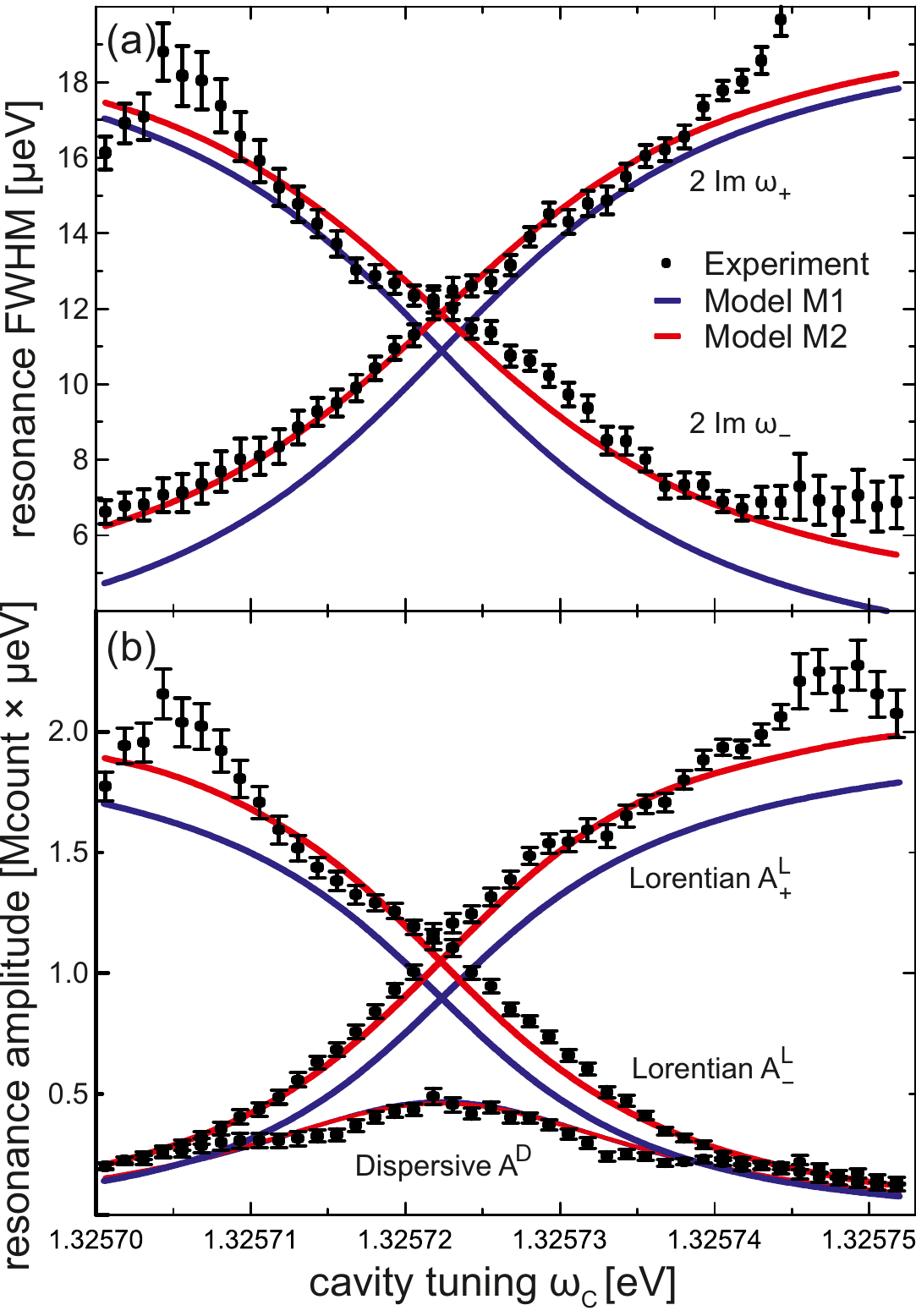} 
\caption{
\label{fig:tunings}
Comparison of model M1 and M2 with the experimental data over the whole cavity--emiter tuning range. (a) The polariton FHWM parameter ($2 \Im \omega_\pm$) and (b) the Lorentzian/dispersive areas $A_\pm^\text{L}$, $A^\text{D}$ versus cavity detuning. M2 provides a much better fit than M1.
}
\end{figure}

As a refinement to the previous model, we incorporate an emitter broadening by convoluting the emitter resonance $\omega_\text{X}$ with a Lorentzian distribution of free FWHM parameter $\Gamma$: this is model M2. The convolution gives an analytical result (see Supplemental Material). A fit of the complete experimental data to this result determines the dynamical parameters ($g,\kappa,\gamma,\epsilon,\Gamma$), as shown in table~\ref{tab:fit}. The model M2 results are shown in figure~\ref{fig:resonance} as the red solid line. The connection to the experimental data is demonstrated also in figure~\ref{fig:tunings}: the parameters $A_\pm^\text{L}$, $A^\text{D}$ and $\omega_\pm$ from eq.~\eqref{eq:decomposition} are shown from both models M1 and M2 along with the experimental data. M2 significantly improves the FWHM parameters $2 \Im \omega_\pm$ and Lorentzian areas $A_\pm^\text{L}$ at all cavity-exciton detuning ranges. Also, M2 resolves the discrepancy in the polariton gap in figure~\ref{fig:resonance}: M2 accounts perfectly for the experimental data both at zero detuning and at large negative detuning. Only M2 is consistent with the experimental data. The microcavity experiment is therefore sensitive to the emitter linewidth in a way that low-power laser spectroscopy alone is not. (We note that the microcavity experiment cannot distinguish easily between a spectral fluctuation and pure exciton dephasing: the M2 predictions are very similar, see Supplemental Material.) The increase in emitter linewidth has a major effect on the cooperativity, table~\ref{tab:fit}: M2 shows that emitter broadening alone reduces $C$ from 9.0, the ``bare" value, to 5.5.

The dynamical parameters of the experiment are summarized in table~\ref{tab:fit}. The freespace emitter lifetime of $800\,\text{ps}$ corresponds to a transform-limited linewidth $\gamma = 0.8\,\mu\text{eV}$ and an dipole moment $\mu_{12} = 1.2\,\text{e}\times\text{nm}$. The microcavity Q-factor $Q = 6\times 10^4$ results in $\kappa = 22\,\mu\text{eV}$. From a simulation of the microcavity, a vacuum electric field maximum of $E_\text{vac} \simeq 2 \times 10^4\,\text{V}/\text{m}$ is expected, yielding $g = \mu_{12} E_\text{vac} \simeq 24\,\mu\text{eV}$. Experimentally, $g$ is smaller than this best-case estimate. From model M1 a cooperativity of $C = 2 g^2 / (\kappa \gamma) = 5.5\pm 0.1$, a result depending only weakly on the model assumptions.

An obvious route to higher cooperativity for the presented microcavity system is to improve the mirrors, i.e.\ to reduce the photon loss rate $\kappa$. Presently, the dielectric DBR is the limiting factor and this can be readily improved with ``supermirror" coatings~\cite{Rempe1992}. The coupling $g$ should also be improved: presently, slight errors in the microcavity manufacture reduce $g$ from its best-case value. However, the point we wish to stress in this work is the equal importance of the emitter dynamics. If the additional broadening can be eliminated by improved emitter quality, the cooperativity can be increased from $C=5.5$ to $C = 9.0$ even without an improvement in the microcavity. This is an entirely realistic proposition: approaches exist by which the additional broadening is routinely sub-$\mu\text{eV}$~\cite{KuhlmannNaturePhysics2013}, in certain cases eliminated altogether~\cite{KuhlmannArxiv2013}, without telegraph-like noise. The present experiment demonstrates that the use of such emitters will easily allow a cooperativity exceeding 10 to be achieved, a powerful route to the application of cavity-QED to quantum control in the solid-state.

\begin{table}
\caption{
\label{tab:fit}
Quantiative fit results of the dynamical parameters for models M1 (no emitter broadening) and for model M2 (with emitter broadening $\Gamma$).
}
\begin{threeparttable}
\begin{ruledtabular}
\begin{tabular}{ c c d d }
Quantity & Unit & \multicolumn{1}{c}{Model M1} & \multicolumn{1}{c}{Model M2} \\
$g$ & $\mu\text{eV}/\hbar$ & 11.05(2) & 11.13(2) \\
$\kappa$ & $\mu\text{eV}/\hbar$ & 19.48(9) & 19.84(9) \\
$\gamma$ & $\mu\text{eV}/\hbar$ & 2.28(4) & 1.38(4) \\
$\Gamma$ & $\mu\text{eV}/\hbar$ & \multicolumn{1}{c}{--} & 1.26(5) \\
$\eta \kappa t \left| \epsilon \right|^2$\tnote{$\ast$} & $\text{Mcount}\,(\mu\text{eV}/\hbar)^2$ & 6.15(4) & 7.08(4) \\
$C = 2 g^2 / (\kappa \gamma) $ & & 5.5(1) & 9.0(3)
\end{tabular}
\end{ruledtabular}
\begin{tablenotes}
\item[$\ast$] with integration time $t = 20\,\text{s}$ and $\eta$ the overall collection efficiency of the cavity emission.
\end{tablenotes}
\end{threeparttable}
\end{table}

\begin{acknowledgments}
We thank P.\ M.\ Petroff for provision of the semiconductor wafer; we acknowledge financial support from NCCR QSIT and from SNF project 200020\_156637. L.G. and S.S. contributed equally to this work.

\end{acknowledgments}



%

\end{document}



\title{Towards high cooperativity strong coupling of a quantum dot in a tunable microcavity: Supplemental Material}



\author{Lukas Greuter}
\author{Sebastian Starosielec}
\author{Andreas V. Kuhlmann}
\author{Richard J. Warburton}
\affiliation{Department of Physics, University of Basel, Klingelbergstrasse 82, Basel 4056, Switzerland}


\date{\today}


\pacs{}

\maketitle 

\section{Sample structure}
\begin{figure}
\includegraphics[width=85mm]{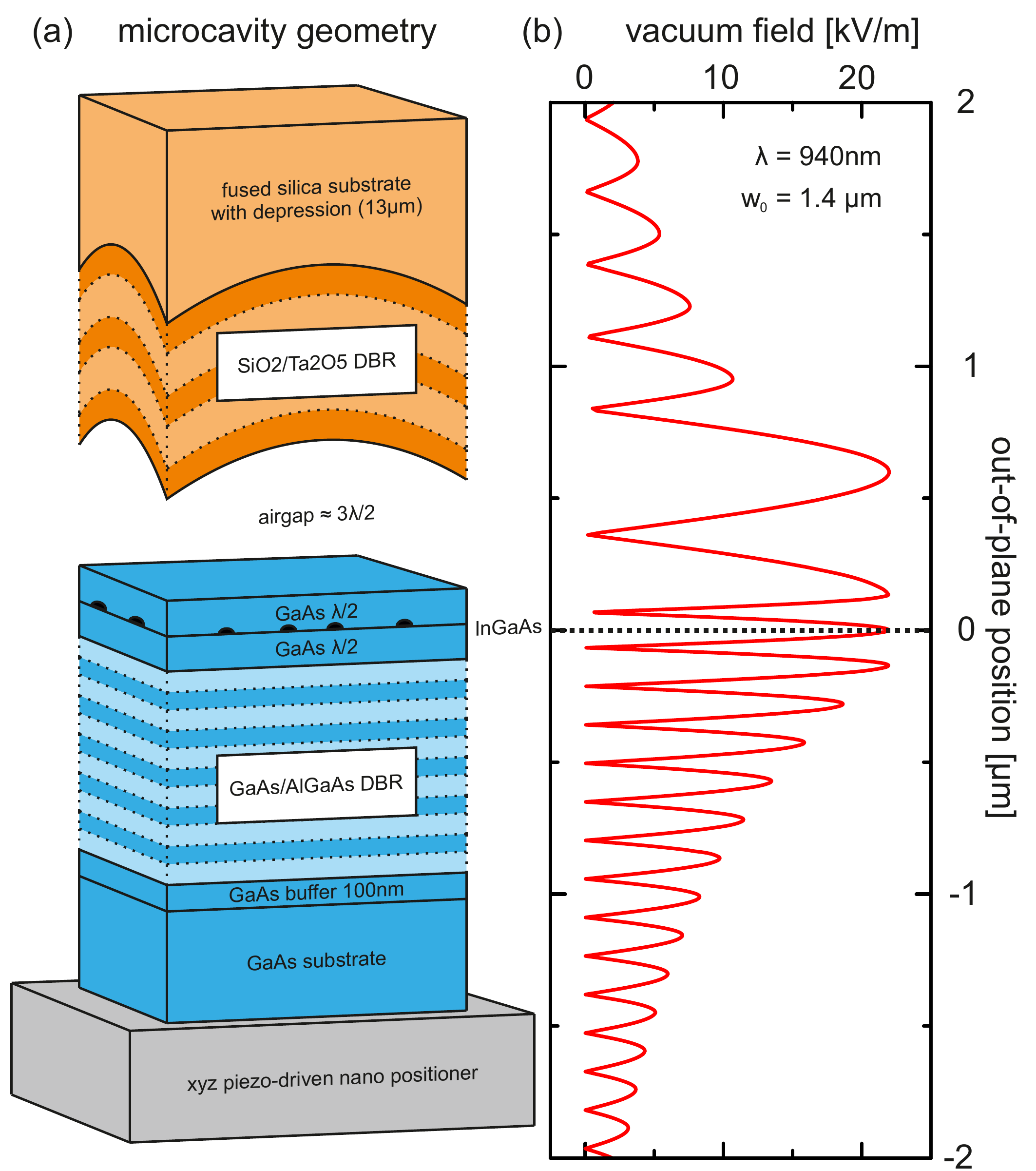} 
\caption{
\label{fig:sample}
(a) Sample structure within cavity configuration and (b) estimated vacuum field distribution for the design wavelength of $\lambda = 940\,\text{nm}$. The field distribution is estimated from one-dimensional transfer matrix methods, with a Gaussian beam waist of $w_0 = 1.4\,\mu\text{m}$.
}
\end{figure}

A self-assembled InGaAs quantum dot is positioned at cavity mode electric field antinode, as depicted in figure~\ref{fig:sample}a. The heterostructure was grown by molecular beam epitaxy by Pierre Petroff at UCSB California, and consists of a $100\,\text{nm}$ GaAs smoothing layer on a GaAs substrate, and a 32.5 pair $\lambda/4$ AlGaAs/GaAs distributed Bragg reflector (DBR) as the bottom mirror of the microcavity, which is terminated by a $\lambda$-layer GaAs host matrix. During growth, the InGaAs wetting layer is inserted at a $\lambda/2$ distance from the sample surface (figure~\ref{fig:sample}b). The top mirror is produced by $\text{CO}_2$ laser ablation from a fused silica substrate, where a concave depression with radius of curvature $\approx 13\,\mu\text{m}$ is created before a $\text{Ta}_2\text{O}_5$ DBR coating is applied by ion-beam sputtering. The nominal reflectivities are $R_\text{bot} = 99.99\%$ and $R_\text{top} = 99.95\%$. The bottom mirror is mounted on an xyz piezo-driven positioner for sub-nm positioning, allowing both spectral and spatial tuning of the microcavity. Estimating from Gaussian optics a beam waist of $w = 1.4\,\mu\text{m}$ at the quantum dot position from the cavity geometry, with one-dimensional transfer matrix method calculations we estimate a vacuum electric field of $E_\text{vac} \approx 2 \times 10^4 \text{V}/\text{m}$. At $4\,\text{K}$, single quantum dots can be addressed in the wavelength range of $930\ldots 960\,\text{nm}$.

\section{Model calculation}

The model Hamiltonian of the main article reads, in the rotating frame 
of the coherent excitation at frequency $\omega_\text{R}$,
\begin{multline}
\label{eq:HamiltonianCavity}
\mathcal{H} = \hbar (\omega_\text{C} - \omega_\text{R}) \, a^\dagger a + \hbar (\omega_\text{X} - \omega_\text{R}) \, b^\dagger b \\
+ \hbar g\, (a^\dagger b + b^\dagger a) + \hbar \epsilon\, (a^\dagger + a)\;,
\end{multline}
where $a$ denotes the bosonic annihilation operator of the cavity (C) and $b$ the fermionic annihilation operator of the exciton transistion (X). Here, $g$ is the coherent cavity-exciton coupling rate, and $\epsilon$ is the coherent excitation rate driving the bare cavity resonance from an external laser field whose linewidth is neglected. 
Treating $\epsilon$ as a perturbation parameter, in the absence of other pumping mechanisms the resulting field amplitudes will be of order $a,b \propto \epsilon$.

The coherent and incoherent evolution of the density matrix $\rho$ is given by the Lindblad operator description
\begin{equation}
\label{eq:rhoevolution}
\begin{split}
\frac{\text{d}\rho}{\text{d}t} = \frac{i}{\hbar} [\rho,\mathcal{H}] 
&+ \frac{\kappa}{2} (2 a \rho a^\dagger - a^\dagger a \rho - \rho a^\dagger a) \\
&+ \frac{\gamma_\text{g}}{2} (2 b \rho b^\dagger - b^\dagger b \rho - \rho b^\dagger b) \\
&+ \frac{\gamma_\text{pd}}{4} (b_z \rho b_z - \rho) \;,
\end{split}
\end{equation}
with the cavity photon loss rate $\kappa$ of the single cavity mode under consideration; $\gamma_\text{g}$ denotes the exciton's spontaneous emission rate into other guided modes of the cavity. For completion, we also consider an exciton pure dephasing contribution $\gamma_\text{pd}$ (where $b_z = 1 - 2 b^\dagger b$), whose effect on the dynamics is considered further below.

Observables $O$ inherit a time-dependent expectation value $\langle O \rangle(t) = \Tr [\rho(t) O]$ from the density matrix. The expectation values of the lowest orders of normal-ordered field operators yield a set of optical Bloch equations
\begin{subequations}
\label{eq:bloch}
\begin{align}
\frac{\text{d}}{\text{d}t} \langle a^\dagger \rangle
&= \left[i (\omega_\text{C} - \omega_\text{R}) - \frac{\kappa}{2} \right] \langle a^\dagger \rangle + i g \langle b^\dagger \rangle + i \epsilon \\
\frac{\text{d}}{\text{d}t} \langle b^\dagger \rangle
&= \left[i (\omega_\text{X} - \omega_\text{R}) - \frac{\gamma_\text{g} + \gamma_\text{pd}}{2} \right] \langle b^\dagger \rangle + i g \langle a^\dagger \rangle \notag \\
& - 2 i g \langle a^\dagger b^\dagger b \rangle \\
\frac{\text{d}}{\text{d}t} \langle a^\dagger a \rangle
&= -\kappa \langle a^\dagger a \rangle - \left[i g \langle a^\dagger b \rangle + i \epsilon \langle a^\dagger \rangle + \text{h.c.} \right] \\
\frac{\text{d}}{\text{d}t} \langle b^\dagger b \rangle
&= -\gamma_\text{g} \langle b^\dagger b \rangle + \left[ i g \langle a^\dagger b \rangle + \text{h.c.} \right] \\
\frac{\text{d}}{\text{d}t} \langle b^\dagger a \rangle 
&= \left[ i (\omega_\text{X} - \omega_\text{C}) - \frac{\gamma_\text{g} + \gamma_\text{pd} + \kappa}{2} \right] \langle b^\dagger a \rangle \notag \\ 
& + i g (\langle a^\dagger a \rangle - \langle b^\dagger b \rangle) - i \epsilon \langle b^\dagger \rangle - 2 i g \langle a^\dagger b^\dagger a b \rangle \;.
\end{align}
\end{subequations}
The higher-order terms $\langle a^\dagger b^\dagger b \rangle$ and $\langle a^\dagger b^\dagger a b\rangle$ originate from the fermionic nature of the exciton after applying the commutator rule $[b,b^\dagger] = 1 - 2 b^\dagger b$ and thus represent all saturation effects. At weak excitations $\epsilon \propto b \ll 1$ these contributions are suppressed and are futher neglected. For vanishing pure dephasing rate $\gamma_\text{pd} \ll \gamma_\text{g}$, the set of optical Bloch equations are solved by the ansatz $\langle a^\dagger a\rangle = \langle a^\dagger \rangle \langle a \rangle$, $\langle b^\dagger b\rangle = \langle b^\dagger \rangle \langle b \rangle$ and $\langle b^\dagger a\rangle = \langle b^\dagger \rangle \langle a \rangle$ with $\langle a^\dagger \rangle$ and $\langle b^\dagger \rangle$ the solution to eq.~(\ref{eq:bloch}a-b). The steady state ($\text{d}/\text{d}t \equiv 0$) yields
\begin{subequations}
\begin{align}
\label{eq:solutiona}
\langle a^\dagger \rangle &= \frac{\epsilon (\omega_\text{X} - \omega_\text{R} + i \tfrac{\gamma_\text{g}}{2})}{g^2 - (\omega_\text{X} - \omega_\text{R} + i \tfrac{\gamma_\text{g}}{2}) (\omega_\text{C} - \omega_\text{R} + i \tfrac{\kappa}{2})}  \\
&= \frac{\epsilon^a_+}{\omega_\text{R} - \omega_+} + \frac{\epsilon^a_-}{\omega_\text{R} - \omega_-} \\
\langle b^\dagger \rangle &= \frac{\epsilon g}{g^2 - (\omega_\text{X} - \omega_\text{R} + i \tfrac{\gamma_\text{g}}{2}) (\omega_\text{C} - \omega_\text{R} + i \tfrac{\kappa}{2})}  \\
&= \frac{\epsilon^b_+}{\omega_\text{R} - \omega_+} + \frac{\epsilon^b_-}{\omega_\text{R} - \omega_-} \;.
\end{align}
\end{subequations}
As a function of the resonant probe $\omega_\text{R}$, a double pole structure arises at complex Rabi frequencies
\begin{equation}
\omega_\pm = \frac{\omega_\text{C} + \omega_\text{X}}{2} + i \frac{\kappa + \gamma_\text{g}}{4} \pm \sqrt{g^2 + \left(\frac{\omega_\text{C} - \omega_\text{X}}{2} + i \frac{\kappa - \gamma_\text{g}}{4} \right)^2}
\end{equation}
with projected excitation rates
\begin{subequations}
\begin{align}
\label{eq:asolution}
\epsilon^a_\pm &= \frac{\epsilon}{2} \left[1 \pm \frac{\tfrac{\omega_\text{C} - \omega_\text{X}}{2} + i \tfrac{\kappa - \gamma_\text{g}}{4}}{\sqrt{g^2 + \left(\tfrac{\omega_\text{C} - \omega_\text{X}}{2} + i \tfrac{\kappa - \gamma_\text{g}}{4} \right)^2}} \right] \\
\epsilon^b_\pm &= \mp \frac{\epsilon}{2} \left[[\frac{g}{\sqrt{g^2 + \left(\tfrac{\omega_\text{C} - \omega_\text{X}}{2} + i \tfrac{\kappa - \gamma_\text{g}}{4} \right)^2}} \right] \;.
\end{align}
\end{subequations}

So far, the detection channel has not been explicitly modeled. A weak coupling of the cavity to a continuum of lossy detection modes contributes a photon flux of $\eta \kappa \langle a^\dagger a \rangle$ to the observed intensity, where the collection efficiency $\eta$ has no dependence on the cavity tuning. In the weak excitation regime, both the absolute value of $\langle a^\dagger a\rangle$ and the excitation rate $\epsilon$ are difficult to determine experimentally. We note that the detected intensity is proportional to $\langle a^\dagger a \rangle$, and limit our study to its dependence on $\omega_\text{R}$. A partial fraction decomposition of the absolute square of $\langle a^\dagger \rangle$ from eq.~\eqref{eq:solutiona} yields
\begin{equation}
\label{eq:aa-lineshape}
\begin{split}
\langle a^\dagger a\rangle &= [V_+ + \Re W] \mathcal{L} (\omega_\text{R} - \omega_+)  + \Im W \mathcal{D} (\omega_\text{R} - \omega_+) \\
&+ [V_- + \Re W] \mathcal{L} (\omega_\text{R} - \omega_-) - \Im W \mathcal{D} (\omega_\text{R} - \omega_-) \;,
\end{split}
\end{equation}
i.e.\ a sum of unit-area Lorentzian and corresponding dispersive function lineshapes
\begin{subequations}
\begin{align}
\mathcal{L}(\omega_\text{R} - \omega_\pm) &= \frac{\Im \omega_\pm / \pi}{(\omega_\text{R} - \Re \omega_\pm)^2 + (\Im \omega_\pm)^2} \\
\mathcal{D}(\omega_\text{R} -  \omega_\pm) &= \frac{(\omega_\text{R} - \Re \omega_\pm) / \pi}{(\omega_\text{R} - \Re \omega_\pm)^2 + (\Im \omega_\pm)^2}
\end{align}
\end{subequations}
with magnitudes
\begin{equation}
\label{eq:amplitudes}
V_\pm = \frac{\pi |\epsilon^a_\pm|^2}{\Im \omega_\pm} \quad \text{and} \quad
W = 2\pi i \frac{\epsilon^a_+ \epsilon^{a\ast}_-}{\omega_+ - \omega_-^\ast}\;,
\end{equation}
where ($\ast$) denotes complex conjugation.
The lineshape resonances are located at $\Re \omega_\pm$ with FHWM parameter $2 \left| \Im \omega_\pm \right|$. 
The result for $\langle b^\dagger b \rangle$ is analogous to eq.\eqref{eq:aa-lineshape}, with $\epsilon^b$ substituted into the magnitudes eq.~\eqref{eq:amplitudes}.

\section{emitter broadening}
In this section, we summarize the effects of two major classes of broadening mechanisms of the exciton on the resonance lineshapes: a pure dephasing, i.e.\ an additional loss of exciton coherence in addition to radiative decay, and a spectral wandering, i.e.\ a temporal fluctuation of the bare exciton transition frequency $\omega_\text{X}$. The dynamics under pure dephasing are governed by the Lindblad operator contribtion proportional to $\gamma_\text{pd}$, the last term in eq.~\eqref{eq:rhoevolution}. We implement the spectral wandering by a convolution of the observable $\langle a^\dagger a \rangle$ with a distribution of $\omega_\text{X}$ with FWHM parameter $\gamma_\text{sw}$. As long as $\gamma_\text{sw}$ is much smaller than the observed linewidths $\approx \kappa$, the details of the distribution shape are insignificant. For the sake of analytical simplicity, we choose a Lorentzian distribution. 

\subsection{Pure dephasing}
The optical Bloch equations eq.~\eqref{eq:bloch} can be solved analytically for a nonzero pure dephasing rate $\gamma_\text{pd}$ within the weak excitation regime. The $\omega_\text{R}$ dependence of the result is
\begin{equation}
\langle a^\dagger a \rangle = \langle a^\dagger a \rangle' + \frac{\mathcal{C}_\text{pd}}{\left| \omega_\text{R} - \omega_+' \right|^2 \left|\omega_\text{R} - \omega_-'  \right|^2}
\end{equation}
where the primed expressions correspond to the previous results when $\gamma_\text{g}$ is renormalized by $\gamma_\text{g} \rightarrow \gamma_\text{g} + \gamma_\text{pd}$. The correction contribution $\mathcal{C}_\text{pd}$ is given by
\begin{equation}
\begin{split}
\mathcal{C}_\text{pd} &= 4 |\epsilon|^2 g^4 \frac{\gamma_\text{pd}}{\gamma_\text{g}} \frac{\kappa + \gamma_\text{g} + \gamma_\text{pd}}{\kappa} \Bigl[ 4g^2 \tfrac{(\kappa + \gamma_\text{g})(\kappa + \gamma_\text{g} + \gamma_\text{dp})}{\kappa \gamma_\text{g}} \\
&+ (\kappa + \gamma_\text{g} + \gamma_\text{dp})^2 + 4(\omega_\text{C} - \omega_\text{X})^2\Bigr]^{-1} \;.
\end{split}
\end{equation}
In the experimental regime of the main article ($g\approx 10\,\mu\text{eV}$, $\kappa \approx 20\,\mu\text{eV}$, $\gamma_\text{g} \approx 2\,\mu\text{eV}$) we expect only a weak dependence of $\mathcal{C}_\text{pd}$ on the experimental control parameters, namely the cavity detuning $\omega_\text{C} - \omega_\text{X}$.

\subsection{Spectral wandering}
The Lorentzian convolution ($\ast$) of $\langle a^\dagger a \rangle$, eq.~\eqref{eq:aa-lineshape}, with respect to $\omega_\text{X}$ with FWHM parameter $\gamma_\text{sw}$ is based on the algebraic form of eq.(\ref{eq:solutiona}). Observing the identity
\begin{equation}
\label{eq:appendix2}
\left| \frac{\omega_\text{X} - A}{\omega_\text{X} - B} \right|^2 \ast \mathcal{L}_\text{sw} = \left| \frac{\omega_\text{X} - A'}{\omega_\text{X} - B'} \right|^2 - \frac{\pi \gamma_\text{sw}}{4} \frac{\left| A - B \right|^2}{\Im B \Im B'} \; \mathcal{L}_{B'}(\omega_\text{X})
\end{equation}
valid in the regime $\Im A, \Im B < 0$, we identify $A = \omega_\text{R} - i \gamma_\text{g}/2$ and $B = A + g^2 / (\omega_\text{C} - \omega_\text{R} + i \kappa/2)$. The primed expressions are renormalized according to $\gamma_\text{g} \rightarrow \gamma_\text{g} + \gamma_\text{sw}$. Here, $\mathcal{L}_{B'}$ is a Lorentzian located at $\Re B'$ with FWHM parameter $2 \Im B'$.
Similar to the pure-dephasing case, we find a corresponding algebraic structure
\begin{equation}
\langle a^\dagger a \rangle = \langle a^\dagger a \rangle' + \frac{\mathcal{C}_\text{sw}}{\left| \omega_\text{R} - \omega_+' \right|^2 \left|\omega_\text{R} - \omega'_-  \right|^2}
\end{equation}
with the correction amplitude from spectral wandering
\begin{equation}
\begin{split}
\mathcal{C}_\text{sw} &= 4 |\epsilon|^2 g^4 \frac{\gamma_\text{sw}}{\gamma_\text{g}} \Bigl[ 4g^2 \frac{\kappa}{\gamma_\text{g}} + \kappa^2 + 4 (\omega_\text{R} - \omega_\text{C})^2\Bigr]^{-1} \;.
\end{split}
\end{equation}
Different to the pure dephasing case, the correction amplitude for spectral wandering $\mathcal{C}_\text{sw}$ depends on $\omega_\text{R} - \omega_\text{C}$. However, as for $\mathcal{C}_\text{pd}$, the dependence on experimental parameters $(\omega_\text{R}, \omega_\text{C})$ is only weak as $g \approx \kappa \gg \gamma_\text{g}$.

\subsection{Lineshape modification}
Treating both correction amplitudes $\mathcal{C}_\text{pd}$, $\mathcal{C}_\text{sw}$ as approximately constant, the emitter broadening induces, along with the renormalization of $\gamma_\text{g}$, a correction to the Lorentzian and dispersive lineshape constituents according to
\begin{equation}
\label{eq:aa-lineshape-pd}
\begin{split}
\langle a^\dagger a\rangle &= \langle a^\dagger a\rangle' \\
&+ \Re U_+ \mathcal{L} (\omega_\text{R} - \omega'_+)  + \Im U_+ \mathcal{D} (\omega_\text{R} - \omega'_+) \\
&+ \Re U_- \mathcal{L} (\omega_\text{R} - \omega'_-) + \Im U_- \mathcal{D} (\omega_\text{R} - \omega'_-) \;,
\end{split}
\end{equation}
with amplitudes
\begin{equation}
U_\pm = \frac{\pi}{\Im \omega'_\pm} \frac{\mathcal{C}}{(\omega'_\pm - \omega'_\mp)(\omega'_\pm - \omega_\mp^{\prime\ast})} \;.
\end{equation}
From symmetry we find $\Im U_+ = - \Im U_-$. In the strong coupling regime, and also for large cavity-emitter detuning, $U_\pm$ is largely real valued. Hence we expect as the main signature of emitter broadening a significant increase of the Lorentzian lineshape contribution, while the dispersive lineshape constituent remains unaffected.

\section{Bare emitter properties}
The previous analysis was limited to the weak excitation regime where a broadening effect on the emitter can be quantified, while the underlying mechanism (pure dephasing or spectral wandering) remained ambiguous. 
This limitation is lifted in the strong excitation regime: when saturation effects become important a distinction can be made. The full cavity-coupled emitter dynamics are difficult to solve, however the bare emitter dynamics are readily accessible.
The bare exciton emission under resonant excitation -- commonly referred to as resonance fluorescence -- follows the Hamiltonian
\begin{equation}
\mathcal{H} = \hbar (\omega_\text{X} - \omega_\text{R}) b^\dagger b + \frac{\hbar \Omega}{2} (b^\dagger + b)\;,
\end{equation}
where $\Omega$ is the Rabi frequency of the resonant excitation of the emitter. As before, we introduce the radiative decay rate $\gamma$ in freespace and pure dephasing rate $\gamma_\text{pd}$ by Lindblad operators. The optical Bloch equations on the exciton population and coherence then read
\begin{subequations}
\begin{align}
\frac{\text{d}}{\text{d}t} \langle b^\dagger b \rangle
&= -\gamma \langle b^\dagger b \rangle - \frac{i \Omega}{2} \langle b^\dagger \rangle + \frac{i \Omega}{2} \langle b \rangle \\
\frac{\text{d}}{\text{d}t} \langle b^\dagger \rangle
&= \left[i (\omega_\text{X} - \omega_\text{R}) - \frac{\gamma + \gamma_\text{dp}}{2} \right] \langle b^\dagger \rangle + \frac{i \Omega}{2} - i \Omega \langle b^\dagger b \rangle \;.
\end{align}
\end{subequations}
The steady-state population results in a Lorentzian line
\begin{equation}
\langle b^\dagger b \rangle = \frac{\Omega^2 \bar\gamma / \gamma}{4(\omega_\text{R} - \omega_\text{X})^2 + \bar\gamma^2 + 2 \Omega^2 \bar\gamma /\gamma}\;,
\end{equation}
with the combined rate $\bar\gamma = \gamma + \gamma_\text{pd}$.
The observed experimental linewidth $\Gamma$, when the emitter is subject to an additional broadening due to spectral wandering $\gamma_\text{sw}$, is after Lorentzian convolution
\begin{equation}
\Gamma = \sqrt{\bar\gamma^2 + 2 \Omega^2 \bar\gamma / \gamma} + \gamma_\text{sw}\;.
\end{equation}
The resonance fluorescence peak intensity $I = \beta \langle b^\dagger b \rangle$ at resonance $\omega_\text{R} = \omega_\text{X}$ is given by 
\begin{subequations}
\label{eq:IRF}
\begin{align}
I &= \beta \frac{\Omega^2}{\bar\gamma \gamma + 2 \Omega^2} \times \frac{\Gamma - \gamma_\text{sw}}{\Gamma} \\
&= I_\text{sat} \left( 1 - \left[\frac{\Gamma_0 - \gamma_\text{sw}}{\Gamma - \gamma_\text{sw}} \right]^2 \right) \;,
\end{align}
\end{subequations}
where $I_\text{sat}$ is the peak intensity at saturation for $\Omega \gg \gamma$, $\Gamma_0 = \gamma + \gamma_\text{pd} + \gamma_\text{sw}$ is the linewidth $\Omega \rightarrow 0$, and $\beta$ is the overall instrumentation factor. 
Equation~(\ref{eq:IRF}b) expresses the power-dependent resonance fluorescence intensity $I$ in terms of convenient observables $I_\text{sat}$ and $\Gamma$, where $\beta$ and the Rabi frequency $\Omega$ have been eliminated. In the case $\gamma_\text{sw} = 0$, the intensity $I$ yields a linear relation to $\Gamma^{-2}$ with intersects at $I_\text{sat}$ and $T_2$-limited rate $\bar\gamma$. A non-vanishing spectral wandering rate $\gamma_\text{sw} \neq 0$ violates the linear relation, allowing $\gamma_\text{sw}$ to be used as a robust fitting parameter.

\subsection{Experiment}
\begin{figure}
\includegraphics[width=85mm]{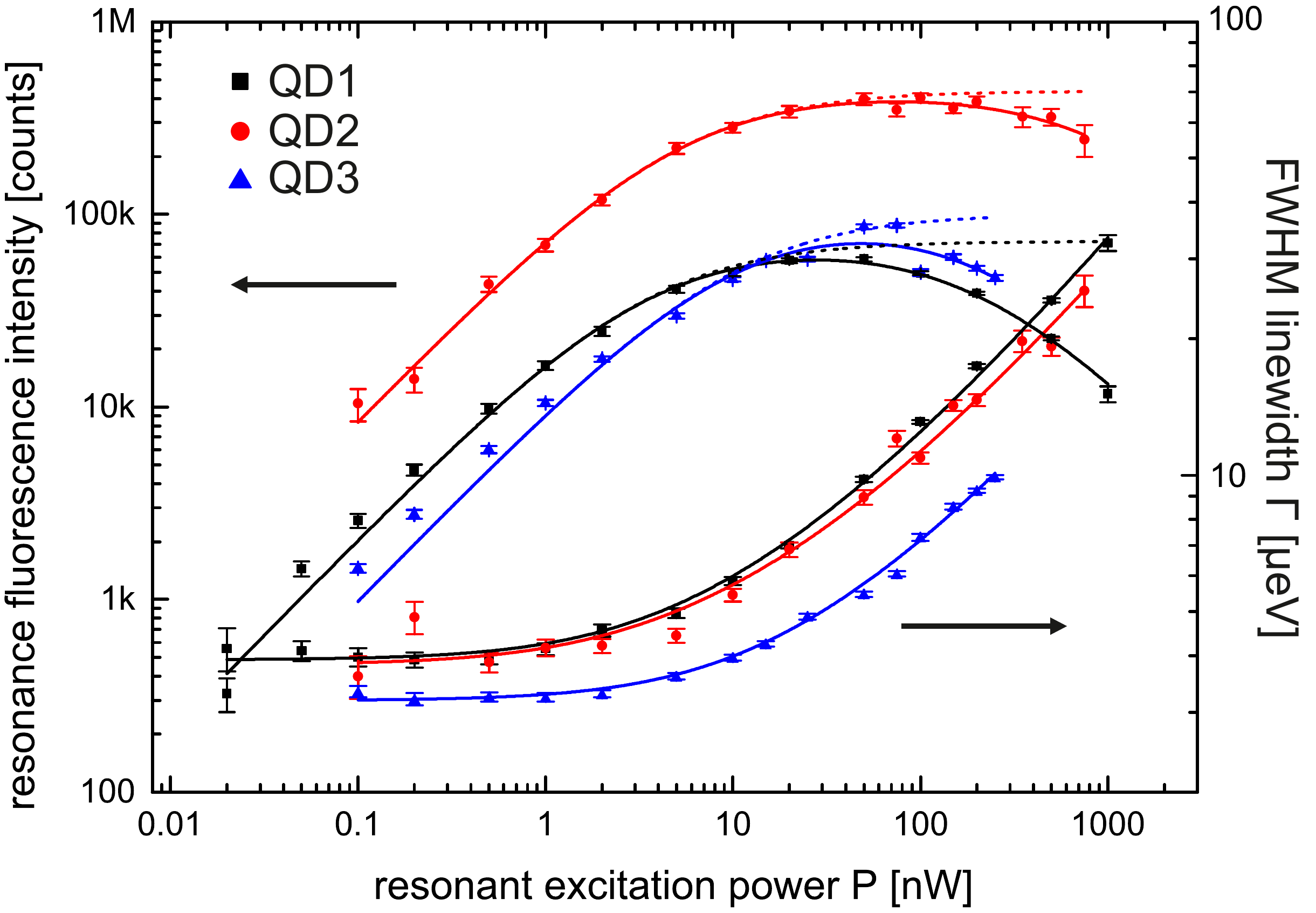} 
\caption{
\label{fig:amplitude}
Resonance fluorescence peak intensity (left scale) and FWHM linewidth (right scale) for three investigated quantum dots (symbols). The peak intensity dependence with resonant pump power matches a three-level-description to a high degree, where the assumed third level is nonresonantly pumped (solid lines). From the three-level description we extrapolate to the corresponding two-level dynamics (dashed lines) where the third level is eliminated from the dynamics. The linewidth dependence with resonant pump power is already well reproduced by the two-level description.
}
\end{figure}

We investigate the spectral wandering of single quantum dots in the same sample area and wavelength as in the microcavity experiment of the main article. Although the very same quantum dot cannot be conserved between configurations, we assume a close statistical resemblance.

\begin{figure}
\includegraphics[width=85mm]{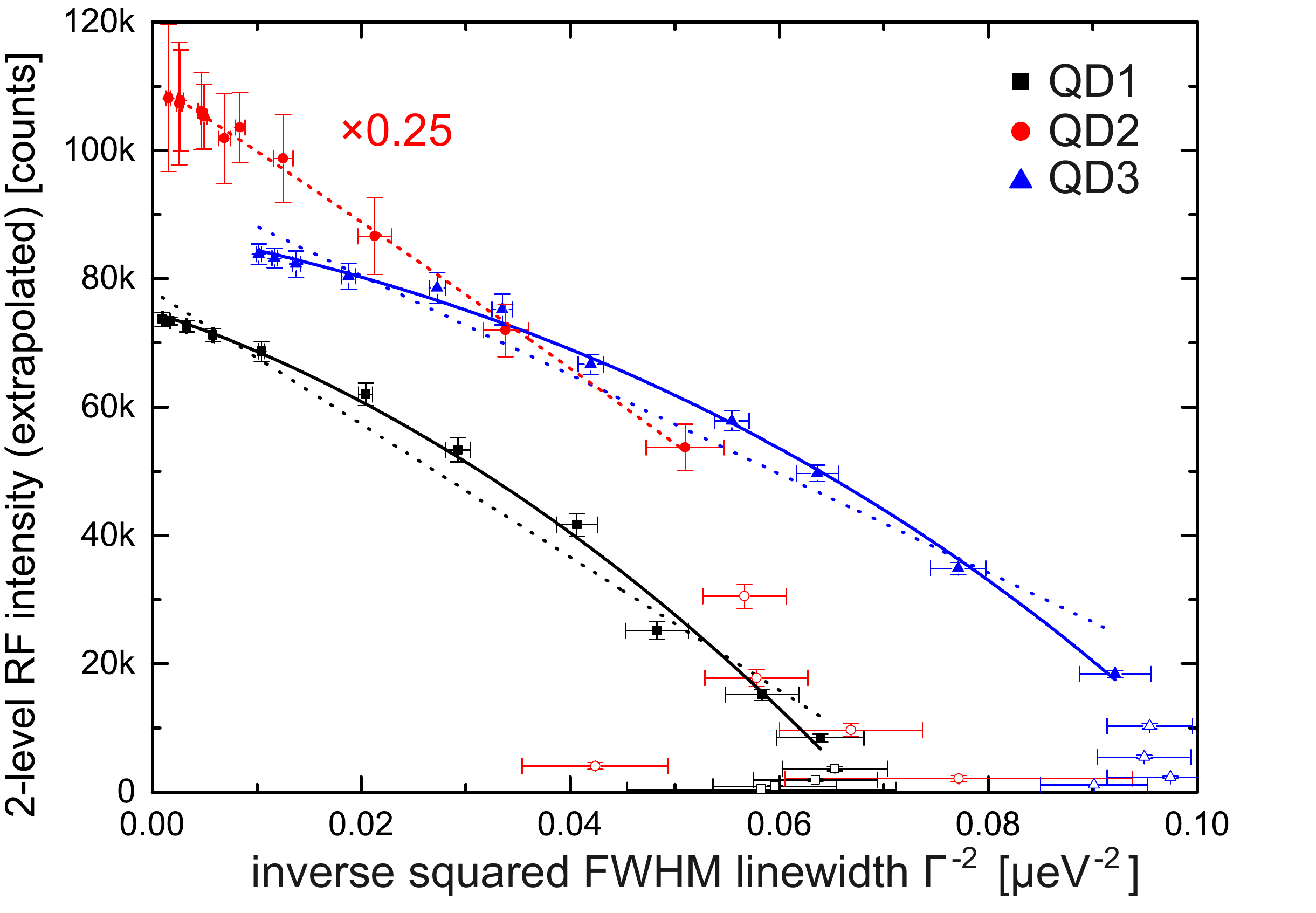}
\caption{
\label{fig:relation}
Measurement of the resonance fluorescence peak intensity versus the inverse squared linewidth (symbols) for the three investigated quantum dots. A vanishing spectral wandering rate yields a linear relation (dotted line), while the experimental data is consistent with a spectral wandering rate of $\approx 1.5\,\mu\text{eV}$ for $\text{QD}_1$ and $\text{QD}_3$. On $\text{QD}_2$ no consistent determination of the spectral wandering rate is found. The open symbols at very low resonant excitation power have been disregarded from the fit, as the collected intensity is dominated by photoluminescence from an ultraweak non-resonant excitation scheme.
}
\end{figure}

Figure~\ref{fig:amplitude} shows as symbols the peak resonance fluorescence intensity $I$ as a function of the resonant excitation power $P$ for three different quantum dots as well as their corresponding resonance FWHM linewidths. Additional with the resonant excitation, we require an ultraweak non-resonant excitation to observe the resonance fluorescence, as was the case in the experiment in the main article. Beyond saturation at about $10\,\text{nW}$ of monitored resonant excitation power, the resonance fluorescence peak intensity drops with further increase in excitation power, in contrast to the two-level model. We attribute this breakdown to a spurious coupling to a third level (e.g. a different charge state, either of the quantum dot or the environment).
Indeed from a simple rate equation model, where a third state is non-resonantly driven from either the upper or lower level at smaller rate $\epsilon P$, the steady-state population of the upper level is 
\begin{equation}
\label{eq:I3level}
I_3 = \beta \frac{(1 + \epsilon \eta_1) P}{\xi_0 + (2 + \epsilon \xi_1) P + \epsilon \xi_2 P^2} \;,
\end{equation}
where the coefficients $\eta_1 < 1$ and $\xi_i$ depend on the details of the relaxation rates. The power dependence of $I_3$ in eq.~\eqref{eq:I3level} is quantitatively well reproduced in the experimental data. Under the assumption $\epsilon \eta_1, \epsilon \xi_1 \ll 1$ we determine $\xi_0$ and $(\epsilon \xi_2)^{-1}$ (see Table~\ref{tab:QD123}). Taking the limit $\epsilon \xi_2 \rightarrow 0$, this allows us to extrapolate from the resonance fluorescence intensity $I_3$ of the three-level system the expected resonance fluorescence intensity $I_2 = P / (\xi_0 + 2P)$ of an effective two-level system where the third level contribution is eliminated. The extrapolated intensity is shown in figure~\ref{fig:amplitude} as dashed line. In terms of resonance linewidth, the experimental data show no significant deviation from a two-level description. 

\begin{table}
\caption{
\label{tab:QD123}
Experimental results on the bare emitter system for $\text{QD}_{1-3}$.
}
\begin{threeparttable}
\begin{ruledtabular}
\newcolumntype{d}{D{.}{.}{5}}
\begin{tabular}{ c c d d d }
Quantity & Unit & \multicolumn{1}{c}{$\text{QD}_1$} & \multicolumn{1}{c}{$\text{QD}_2$} & \multicolumn{1}{c}{$\text{QD}_3$} \\
$\lambda$ & $\text{nm}$ & 941.79 & 937.41 & 939.04 \\
$\xi_0$ & $\text{nW}$ & 7.0(5) & 10.3(9) & 20.3(65) \\
$(\epsilon \xi_2)^{-1}$ & $\mu\text{W}$ & 0.111(9) & 0.55(8) & 0.113(51) \\
$\Gamma_0$ & $\mu\text{eV}$ & 3.84(4) & 3.17(8) & 3.10(2) \\
$\gamma_\text{sw}$ & $\mu\text{eV}$ & 1.4(3) & 0.2(3) & 1.5(1) \\
$\gamma_\text{pd}$ & $\mu\text{eV}$ & \approx 1.6 & $---$\tnote{$\ast$} & \approx 0.8 \\
\end{tabular}
\end{ruledtabular}
\begin{tablenotes}
\item[$\ast$] No consistent determination of $\gamma_\text{sw}$ was found for $\text{QD}_2$.
\end{tablenotes}
\end{threeparttable}
\end{table}

Figure~\ref{fig:relation} shows as symbols the resonance fluorescence intensity as a function of the inverse squared linewidth $\Gamma^{-2}$ for the three investigated quantum dots (filled symbols). At very low resonant excitation powers, the collected intensity is dominated by the photoluminescence intensity from the additional ultraweak non-resonant excitation scheme. For this reason, we discard the data for very low collected intensities (open symbols). Applying the relation eq.~(\ref{eq:IRF}b) to the data, for $\text{QD}_1$ and $\text{QD}_3$, the relation is well reproduced for $\gamma_\text{sw} = 1.5\pm 0.1\,\mu\text{eV}$ and $1.4\pm 0.2\,\mu\text{eV}$ respectively (solid line). For comparison the best fit for $\gamma_\text{sw} = 0$ (dotted line) is in clear contradiction to the experimental data. For $\text{QD}_2$ no significant spectral wandering is observed, however we note that the relative error on the resonance fluorescence intensity is considerably larger than for the other QDs and no consistent behaviour at low intensity is found. Thus on $\text{QD}_2$ no reliable estimation of the spectral wandering rate can be obtained. 
The $T_2$-limited linewidth $\bar\gamma = \gamma + \gamma_\text{pd} = \Gamma_0 - \gamma_\text{sw}$ evaluates to $\approx 2.44\,\mu\text{eV}$ ($1.6\,\mu\text{eV}$) for $\text{QD}_1$ ($\text{QD}_3$). As the transform-limited radiative decay rate $\gamma \approx 0.8\,\mu\text{eV}$, we estimate a corresponding pure dephasing rate of $\gamma_\text{pd} \approx 1.6\,\mu\text{eV}$ ($\approx 0.8\,\mu\text{eV}$) for $\text{QD}_1$ ($\text{QD}_3$).

In summary, we observe that spectral wandering is likely to represent dominating broadening mechanism in the investigated sample. This result underlines the major statement of the main article: the cavity-coupled exciton cooperativity can be readily enhanced if the additional emitter broadening, identified as spectral wandering, can be reduced.

\begin{acknowledgments}
\end{acknowledgments}

\bibliography{references0}